# K-sort: A new sorting algorithm that beats Heap sort for n ≤ 70 lakhs!

Kiran Kumar Sundararajan[1], Mita Pal[2], Soubhik Chakraborty[3*] and N.C. Mahanti[4]

[1]Barclays Bank PLC, United Arab Emirates, Dubai

[2-4]Department of Applied Mathematics, Birla Institute of Technology, Mesra, Ranchi-835215, India

*corresponding author's email: soubhikc@yahoo.co.in

Abstract:

Sundararajan and Chakraborty [10] introduced a new version of Quick sort removing the interchanges. Khreisat [1] found this algorithm to be competing well with some other versions of Quick sort. However, it uses an auxiliary array thereby increasing the space complexity. Here, we provide a second version of our new sort where we have removed the auxiliary array. This second improved version of the algorithm, which we call K-sort, is found to sort elements faster than Heap sort for an appreciably large array size (n ≤ 70,00,000) for uniform U[0, 1] inputs.

**1. Introduction:**

There are several internal sorting methods (where all the sorting elements can be kept in the main memory). The simplest algorithms like bubble sort usually takes $O(n^2)$ time to sort n objects and are only useful for sorting short lists. One of the most popular sorting algorithms for sorting long lists is Quick sort, which takes $O(n\log_2 n)$ time on an average and $O(n^2)$ in the worst case. For a comprehensive literature on sorting algorithms, see Knuth [2].

Sundararajan and Chakraborty [10] introduced a new version of Quick sort removing the interchanges. Khreisat [1] found this algorithm to be competing well with some other versions of Quick sort like SedgewickFast, Bsort and Singleton sort for n between 3000 to 200,000. Since comparisons and not interchanges are dominant in sorting, the removal of interchanges does not make the order of complexity of this algorithm differ from that of Quick sort. In other words, the algorithm has average and worst case complexity similar to Quick sort, that is, $O(n\log_2 n)$ and $O(n^2)$ respectively which is also confirmed by Khreisat [1]. However, it uses an auxiliary array thereby increasing the space complexity. Here, we provide a second improved version of our new sort, which we call K-sort, where we have removed the auxiliary array. K-sort is found to sort elements faster than Heap sort for an appreciably large array size (n ≤ 70,00,000) for uniform U[0, 1] inputs.

## 1.1 K-sort:

The steps of K-sort are given below:-

Step-1: Initialize the first element of the array as the key element and i as left, j as (right+1), k = p where p is (left+1).

Step-2: Repeat step-3 till the condition (j-i) ≥ 2 is satisfied.

Step-3: Compare a[p] and key element. If  key ≤ a[p]

    then

     Step-3.1:  if ( p is not equal to j and j is not equal to (right + 1) )

         then set a[j] = a[p]

         else if  ( j equals (right + 1)) then

             set temp = a[p] and flag = 1

         decrease j by 1 and assign p = j

    else (if the comparison of step-3 is not satisfied i.e. if key > a[p] )

    Step-3.2: assign a[i] = a[p] , increase i and k by 1  and set p = k

Step-4: set a[i] = key

    if (flag = = 1) then

       assign a[i+1] = temp

Step-5: if ( left < i - 1 ) then

       Split the array into sub array from start to i-th element and repeat steps 1-4 with the sub array

Step-6: if ( left > i + 1 ) then

Split the array into sub array from i-th element to end element and repeat steps 1-4 with the sub array

## 1.2 Illustration:

| | | | | | | | | | | |
|---|---|---|---|---|---|---|---|---|---|---|
| Unsorted List | 55 | 66 | 60 | 78 | 22 | 50 | 75 | 5 | 8 | 94 |
| Key=55 | 8 | 5 | 50 | 22 | **55** | 66 | 75 | 78 | 60 | 94 **Temp = 66** |
| Key=8 | **5** | **8** | 50 | 22 | **55** | 66 | **75** | **78** | **60** | 94 **Temp = 50** |
| Key=50 | **5** | **8** | **22** | **50** | **55** | 66 | **75** | **78** | **60** | 94 **Temp =Nil** |
| Key=66 | **5** | **8** | **22** | **50** | **55** | 60 | **66** | 78 | 75 | 94 **Temp = 75** |
| Key=78 | **5** | **8** | **22** | **50** | **55** | **60** | **66** | 75 | **78** | 94 **Temp = 94** |
| Sorted List | **5** | **8** | **22** | **50** | **55** | **60** | **66** | **75** | **78** | **94** |

Note: If the sub array has single value it need not be processed.

## 2. Empirical (average case time complexity) results:

A computer experiment is a series of runs of a code for various inputs (see Sacks et. al. [9]). By running computer experiments on **Borland International Turbo 'C++' ver 5.02**, we could compare the average sorting time in sec (average taken over 500 readings) for different values of n for both K-sort and Heap sort. Using Monte Carlo simulation (see Kennedy and Gentle [7]), the array of size n was filled with independent continuous uniform U[0, 1] variates and the elements are copied to another array. One array is sorted by K-sort while the other is sorted by Heap sort. Table 1 and fig. 1 gives the empirical results.

**Table1: Average sorting time comparison:**

| n | $n\log_2(n)$ | K- sort avg time (in sec) | Heap sort avg time (in sec) |
|---|---|---|---|
| 100000 | 1660964.05 | 0.0157 | 0.0156 |
| 500000 | 9465784.28 | 0.0811 | 0.1061 |
| 1000000 | 19931568.6 | 0.1877 | 0.2751 |
| 2500000 | 53133741.7 | 0.6532 | 0.8953 |
| 5000000 | 111267483 | 1.914 | 2.1640 |
| 6000000 | 135099186 | 2.5967 | 2.7235 |
| 7000000 | 159172464 | 3.2892 | 3.3749 |
| 7100000 | 161591652 | 3.4502 | 3.3782 |
| 7500000 | 171288444 | 4.0695 | 3.6439 |
| 10000000 | 232534967 | 8.2293 | 5.5951 |

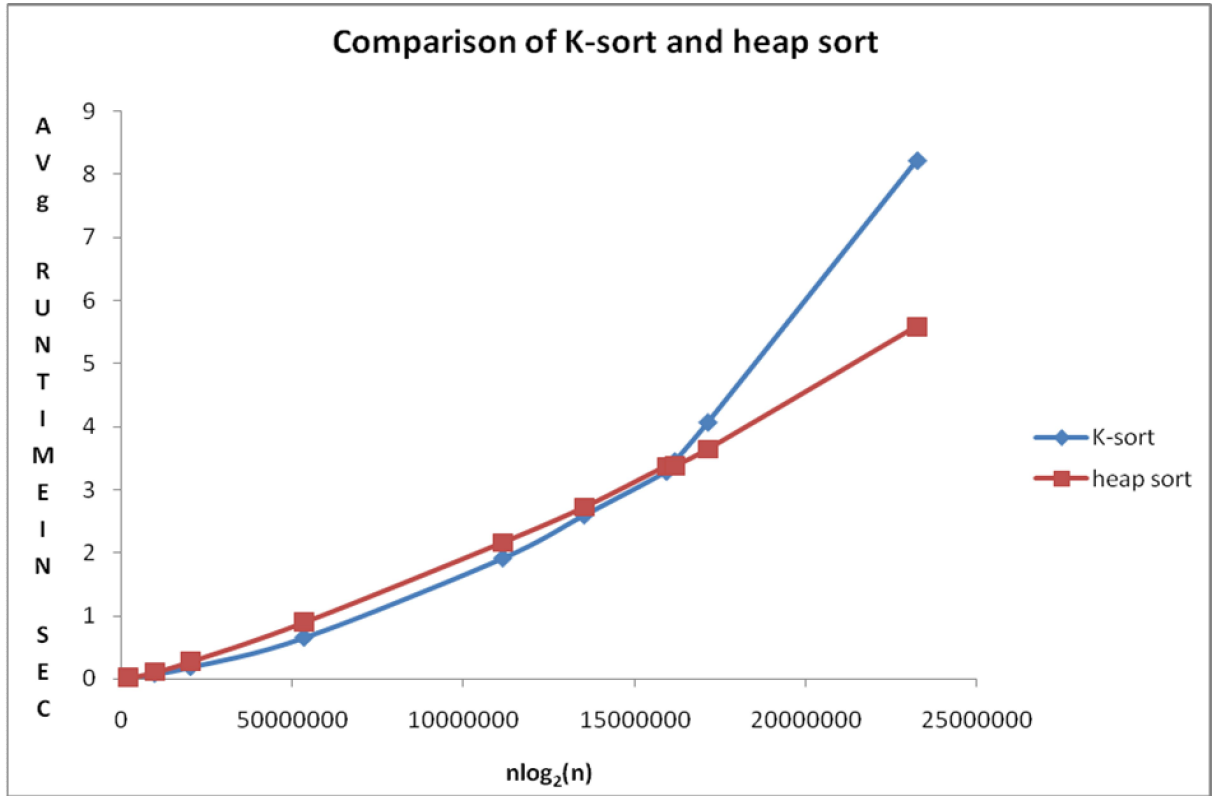

**Fig-1: Graph of K- sort and Heap sort**

The observed average times from continuous uniform distribution U(0,1) input for K-sort and Heap sort are depicted in table-1. Figure-1 together with table 1 suggests a comparison between these algorithms.

A moment's reflection from table 1 suggests that the average run time for K-sort is less than that of Heap sort when the array size n ≤ 70 lakhs and above this range Heap sort is faster.

### 3. Statistical analysis (using Minitab version 15) of the empirical results

### 3.1. Analysis for K-sort: Regressing average sorting time y(K) over nlog $_2$(n) and n

```
The regression equation is
Y(K) = 0.7516 + 0.00000048 nlog₂(n) - 0.00001048 n     ..................................(1)

Predictor            Coef      SE Coef        T      P       VIF
Constant           0.7516       0.4153     1.81  0.113
nlog(n)        0.00000048   0.00000010     4.89  0.002  2225.579
n             -0.00001048   0.00000229    -4.58  0.003  2225.579

S = 0.499133    R-Sq = 97.0%    R-Sq(adj) = 96.1%

PRESS = 8.44451    R-Sq(pred) = 85.42%
```

```
Analysis of Variance

Source          DF     SS       MS        F        P
Regression       2   56.177   28.088   112.74   0.000
Residual Error   7    1.744    0.249
Total            9   57.921

Source     DF   Seq SS
nlog(n)     1   50.942
n           1    5.235

Obs    nlog₂(n)     y(K)     Fit     SE Fit   Residual   St Resid
 1      1660964    0.016    0.501    0.366    -0.485     -1.43
 2      9465784    0.081    0.054    0.274     0.027      0.06
 3     19931569    0.188   -0.163    0.238     0.350      0.80
 4     53133742    0.653    0.052    0.265     0.602      1.42
 5    111267483    1.914    1.751    0.247     0.163      0.38
 6    135099186    2.597    2.709    0.217    -0.112     -0.25
 7    159172464    3.289    3.782    0.202    -0.493     -1.08
 8    161591652    3.450    3.895    0.202    -0.445     -0.97
 9    171288444    4.069    4.356    0.209    -0.287     -0.63
10    232534967    8.229    7.550    0.422     0.679      2.54R

R denotes an observation with a large standardized residual.
Fig. 2.1-2.4 give a graphical summary of some further tests of model fit.
```

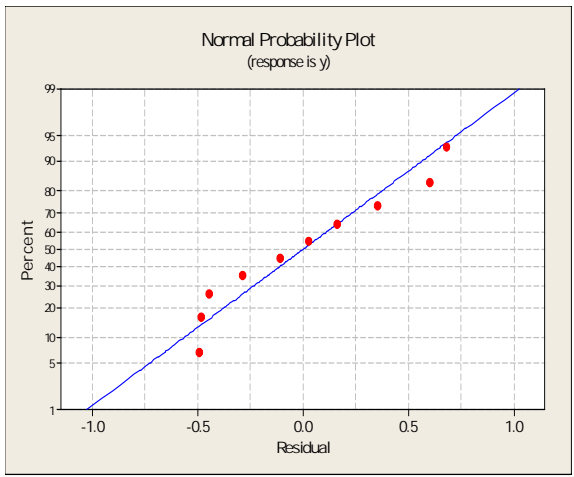
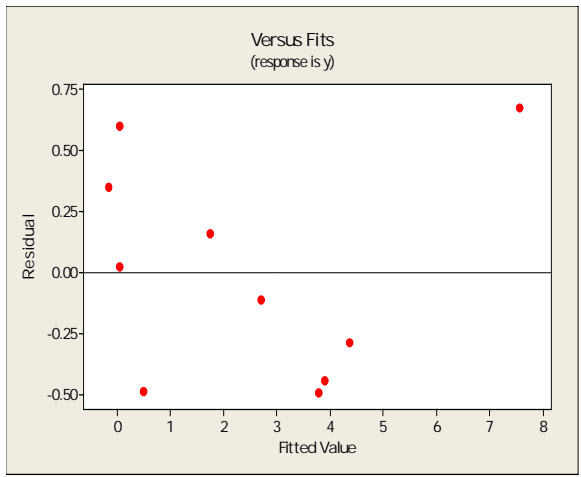

Fig. 2.1 Normal Probability Plot          Fig. 2.2 Residual versus fitted value

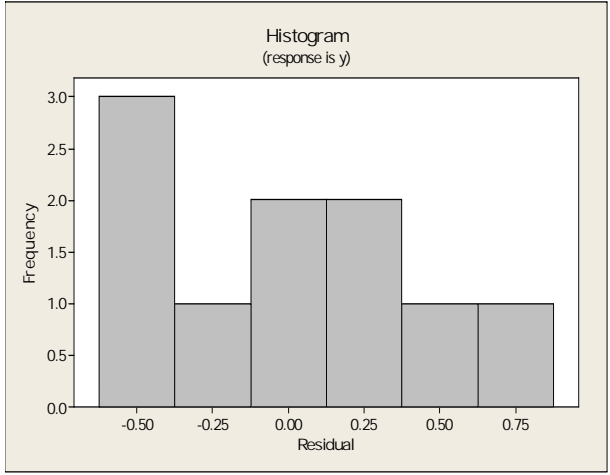 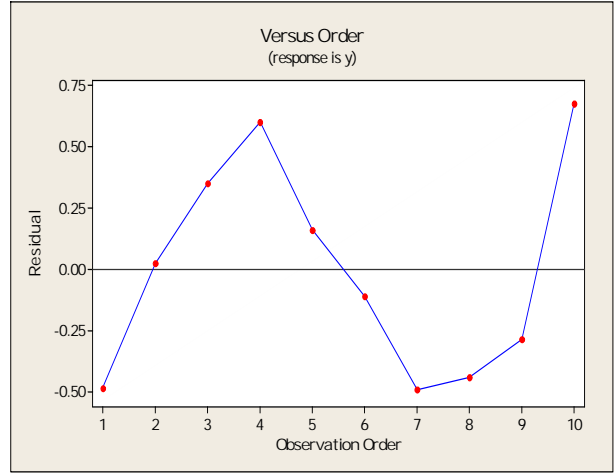

Fig. 2.3 Histogram of residual                Fig. 2.4 Residual versus observation order

## 3.2. Analysis for Heap sort: Regressing average sorting time y(H) over nlog$_2$ (n) and n

```
The regression equation is
Y(H) = 0.12574 + 0.00000013 nlog₂(n) - 0.00000256 n      .........................(2)

Predictor           Coef      SE Coef        T      P      VIF
Constant         0.12574      0.06803     1.85  0.107
nlog(n)       0.00000013   0.00000002     8.29  0.000  2225.579
n            -0.00000256   0.00000037    -6.85  0.000  2225.579

S = 0.0817608    R-Sq = 99.9%    R-Sq(adj) = 99.8%

PRESS = 0.225845    R-Sq(pred) = 99.28%

Analysis of Variance

Source            DF       SS       MS        F      P
Regression         2   31.169   15.585  2331.34  0.000
Residual Error     7    0.047    0.007
Total              9   31.216

Source    DF  Seq SS
nlog(n)    1  30.856
n          1   0.313
```

```
Obs    nlog₂(n)    y(H)      Fit    SE Fit    Residual    St Resid
 1     1660964    0.0156    0.0907   0.0600    -0.0751      -1.35
 2     9465784    0.1061    0.1055   0.0449     0.0006       0.01
 3    19931569    0.2751    0.2186   0.0391     0.0565       0.79
 4    53133742    0.8953    0.7985   0.0434     0.0968       1.40
 5   111267483    2.1640    2.1379   0.0404     0.0261       0.37
 6   135099186    2.7235    2.7507   0.0355    -0.0272      -0.37
 7   159172464    3.3749    3.3958   0.0331    -0.0209      -0.28
 8   161591652    3.3782    3.4618   0.0331    -0.0836      -1.12
 9   171288444    3.6439    3.7289   0.0343    -0.0850      -1.14
10   232534967    5.5951    5.4832   0.0691     0.1119       2.56R
```

R denotes an observation with a large standardized residual.

Fig. 3.1-3.4 give a graphical summary of some further tests of model fit.

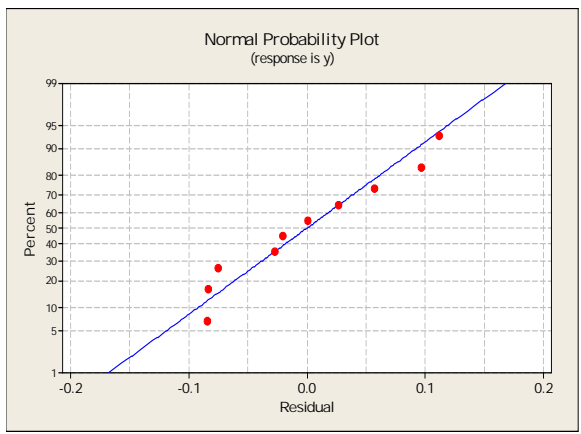

Fig. 3.1 Normal Probability Plot

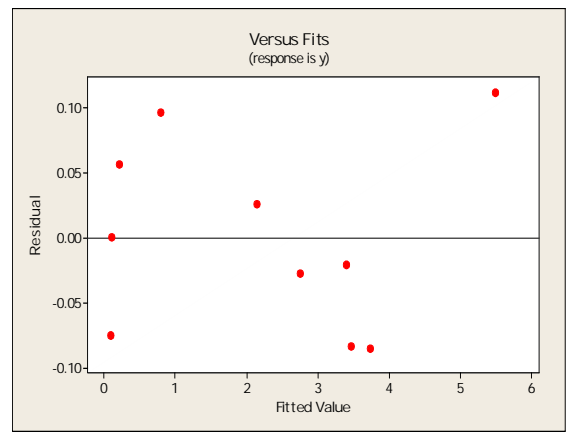

Fig. 3.2 Residual versus fitted value

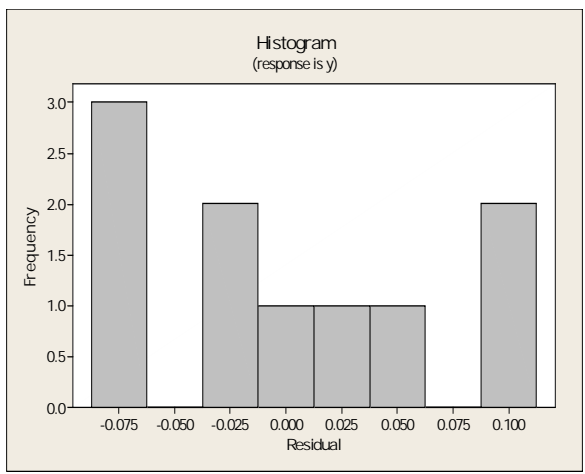

Fig. 3.3 Histogram of residual

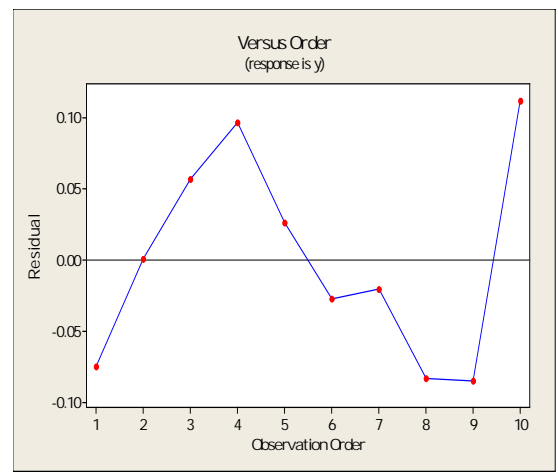

Fig. 3.4 Residual versus observation order

## 4. Discussion

It is easy to see that the sum of squares contributed by $n\log_2 n$ to the regression model, in both K-sort and Heap sort, are substantial in comparison to that contributed by n. Recall that both algorithms have average $O(n\log_2 n)$ complexity. Thus the experimental results are confirming the theory. We kept an n term in the model because a look at the mathematical statement leading to the $O(n\log_2 n)$ complexity in Quick sort and Heap sort does suggest an n term (see Knuth [2]).

The comparing regression equation between the two sorting algorithms for average case is obtained simply by subtraction y(H) from y(K).

We have, y(K)-y(H) = 0.52586 + 0.00000035 $n\log_2(n)$ – 0.00000792 n ……..(3)

The advantage of equations (1), (2) and (3) is that we can predict average run times of both sorting algorithms as well as their difference even for huge values of n for which it may be computationally cumbersome to run the code. Such "cheap prediction" is the motto in computer experiments and permits us to go for stochastic modeling even for non-random data. Another advantage is that knowledge of only the size of the input is enough to make a prediction. That is, the entire input (for which the response is fixed) need not be supplied. Thus prediction through a stochastic model is not only cheap but also more efficient (Sacks et al., [9]).

It is important to note that when we are directly working on program run time, we are actually estimating a *statistical bound* over a finite range (no computer experiment can be performed for infinite input size). A statistical bound differs from a mathematical bound in the sense that unlike a mathematical bound, it *weighs* rather than counts the computing operations and as such it is capable of mixing different operations into a conceptual bound while mathematical complexity bounds are operation specific. Here, time of an operation is taken as its weight. For a general discussion on statistical bound including a formal definition and other properties, see Chakraborty and Sourabh [5]. See also Chakraborty, Modi and Panigrahi, [4] to know why the statistical bound is the ideal bound in parallel computing. The estimate of a statistical bound is obtained by running computer experiments, where the weights are assigned numerical values, over a finite range. This means the credibility of the bound estimate depends on a proper design and analysis of our computer experiment. Further literature on computer experiments with other application areas such as VLSI design, combustion, heat transfer etc. can be found in (Fang, Li and Sudjianto, [3]). See also its review (Chakraborty [6]).

## 5. Conclusion and suggestions for future work:

K-sort is evidently faster than the Heap sort for number of sorting elements up to 70 lakhs, although both the algorithms have same order of complexity $O(n\log_2 n)$ in the average case. Future work involves a study on parameterized complexity (Mahmoud, [8]) on this improved version. As a final comment, we strongly recommend K- sort at least for n ≤ 70, 00000.

However, we agree to opt for Heap-sort in worst case, due to its maintaining $O(n\log_2 n)$ complexity even in the worst case, although it is more difficult to program.

**References :**

[1]. Khreisat, L., *QuickSort A Historical Perspective and Empirical Study*, International Journal of Computer Science and Network Security, VOL.7 No.12, December 2007, p. 54-65

[2] Knuth, D. E., *The Art of Computer Programming* , Vol. 3: Sorting and Searching, Addison Wesely (Pearson Education Reprint), 2000.

[3] Fang, K. T., Li, R. and Sudjianto, A., *Design and Modeling of Computer Experiments* Chapman and Hall, 2006

[4] S. Chakraborty, S., Modi, D. N. and Panigrahi, S., *Will the Weight-based Statistical Bounds Revolutionize the IT?,* International Journal of Computational Cognition, Vol. 7(3), 2009, 16-22

[5] Chakraborty, S. and Sourabh, S. K., *A Computer Experiment Oriented Approach to Algorithmic Complexity*, Lambert Academic Publishing, 2010

[6] Chakraborty, S. Review of the book *Design and Modeling of Computer Experiments* authored by K. T. Fang, R. Li and A. Sudjianto, Chapman and Hall, 2006, published in Computing Reviews, Feb 12, 2008,

http://www.reviews.com/widgets/reviewer.cfm?reviewer_id=123180&count=26

[7] Kennedy, W. and Gentle, J., Statistical Computing, Marcel Dekker Inc., 1980

[8] Mahmoud, H.,*Sorting: A Distribution Theory, John Wiley and Sons*, 2000

[9]. Sacks, J., Weltch, W., Mitchel, T. and Wynn, H., *Design and Analysis of Computer Experiments*, Statistical Science 4 (4), 1989

[10] Sundararajan , K. K. and Chakraborty , S., *A New Sorting Algorithm*, Applied Math. and Compu., Vol. 188( 1), 2007, p. 1037-1041